\documentclass[aps,prl,twocolumn, preprintnumbers]{revtex4}
\usepackage{mathrsfs}
\usepackage{amsmath, amssymb, amsfonts, bbm,slashed, youngtab}
\usepackage{graphics,epsfig,color,subfigure,graphicx,epsf}

\newcommand{\beq}{\begin{equation}}
\newcommand{\eeq}{\end{equation}}
\newcommand{\bea}{\begin{eqnarray}}
\newcommand{\eea}{\end{eqnarray}}

\def\nnb{\nonumber}

\def\nnb{\nonumber}

\newcommand{\nn}{\nonumber}
\newcommand{\benn}{\begin{displaymath}}
\newcommand{\eenn}{\end{displaymath}}

\newcommand{\tr}{{\rm tr}}

\begin{document}

\preprint{UMD-DOE/40762-464}

\title{Color Superconductivity at Large N: A New Hope}

\author{Michael I. Buchoff}
\email{mbuchoff@physics.umd.edu}
\affiliation{Maryland Center for Fundamental Physics,\\ Department of Physics, University of Maryland,
College Park, MD 20742-4111}

\author{Aleksey Cherman}
\email{alekseyc@physics.umd.edu}
\affiliation{Maryland Center for Fundamental Physics,\\ Department of Physics, University of Maryland,
College Park, MD 20742-4111}

\author{Thomas D. Cohen}
\email{cohen@physics.umd.edu}
\affiliation{Maryland Center for Fundamental Physics,\\ Department of Physics, University of Maryland,
College Park, MD 20742-4111}

\begin{abstract}
 At zero density, the `t Hooft large $N_c$ limit often provides some very useful qualitative insights into the non-perturbative physics of QCD.   However, it is known that at high densities the `t Hooft large $N_c$ world looks very different from the $N_c=3$ world, which is believed to be in a color superconducting phase at high densities.  At large $N_c$, on the other hand, the DGR instability causes a chiral-density wave phase to dominate over the color superconducting phase.  There is an alternative large $N_c$ limit, with the quarks transforming in the two-index antisymmetric representation of the gauge group, which at $N_c=3$ reduces to QCD but looks quite different at large $N_c$.  We show that in this alternative large $N_c$ limit, the DGR instability does not occur,  so that it may be plausible that the ground state of high-density quark matter is a color superconductor even when $N_c$ is large.  This revives the hope that a large $N_c$ approximation might be useful for getting some insights into the high-density phenomenology of QCD.
\end{abstract}

\maketitle
\section{Introduction}
The behavior of quark matter at high densities has long been a subject of intense interest.   As a purely theoretical matter, the question of how the system behaves at high densities is of interest as it gives insight into the structure of the theory.  Phenomenologically, the high-density physics of QCD is important for the physics of neutron stars. Unfortunately, there are no systematic and reliable tools to study the properties of QCD matter at phenomenologically relevant densities, when the quark chemical potential $\mu$ is of order the strong scale $\Lambda_{QCD}$.  In contrast, at asymptotically  large densities ($\mu \gg \Lambda_{QCD}$),  resummed perturbation theory becomes a reliable tool for describing instabilities.   It is well known that at asymptotically high densities, an instability develops toward the breaking of  $SU(3)$ color symmetry with the formation of a $\langle q q \rangle$ condensate.   At least in the $SU(3)$ flavor limit, QCD is thus widely believed to be in a color-flavor locked (CFL) phase~\cite{Alford:2007xm}.

There is a hope that color superconductivity may persist to phenomenologically realistic densities.  Unfortunately, there are no known model-independent ways to study the behavior of QCD matter away from the asymptotically high-density domain.  In the region of phenomenological interest,  there is no obvious separation of scales on which to build systematic effective field theories, and the relatively strong coupling prevents the use of perturbative techniques.  Lattice QCD cannot currently be used to study cold finite density systems due to the fermion sign problem, which makes Monte Carlo approaches impractical.   Thus, typically one is forced to use models with varying degrees of connection to QCD to explore the behavior of quark matter at realistic densities.   Given this state of affairs, any approach with a systematic connection to QCD that gives even qualitative insights into moderate density physics would be valuable.  One hope is that large $N_c$ QCD and the $1/N_c$ expansion could be such an approach.

At zero density, the `t Hooft large $N_c$ limit \cite{'tHooft:1973jz, Witten:1979kh} can often provide very useful qualitative insights on some non-perturbative questions; occasionally semi-quantitative predictions can be made ({\it e.g.}, predictions on the size of nucleon mass splittings \cite{Jenkins:1995td,Jenkins:2000mi}).  For this to work, the large $N_c$ world needs to be close to the $N_c=3$ world for the observables of interest.  However, the `t Hooft large $N_c$ world appears to be very different from the $N_c=3$ world at finite density.  In particular, as noted above,  at asymptotically high densities $N_c=3$ QCD has a color-superconducting (`BCS') instability \cite{Son:1998uk}, and is expected to form a color superconductor with a non-trivial $\langle qq \rangle$ condensate.   In contrast, at large $N_c$, at high density  ( $1\ll  \mu/\Lambda_{\mathrm{QCD}} \sim N_c^{0}$ )  there is another instability, first noted by Deryagin, Grigoriev, and Rubakov (DGR),  toward the formation of a chiral density wave, {\it i.e.}, a  $\langle \bar{q}(x) q(y)\rangle$ condensate~\cite{Deryagin:1992rw, Shuster:1999tn,Park:1999bz}.  The DGR condensate dominates over the BCS instability.  Ultimately, the reason for this is that the DGR condensate is a global color singlet (although it breaks gauge invariance), in contrast to the $\langle q q\rangle$ BCS condensate.  So at large $N_c$, high density QCD cannot be in a color superconducting phase; it is either  in a  `chiral density wave'  phase  or in some as yet undiscovered phase.  Thus it might appear that there is little hope that large $N_c$ analyses could have anything useful to say about finite density physics in $N_c=3$ QCD.

However, such a conclusion is  premature as it relies on  the `t Hooft large $N_c$ limit.  The large $N_c$ world is not unique:  there is more than one way to extrapolate from the $N_c=3$ world to large $N_c$.  Different extrapolations yield qualitatively different large $N_c$ worlds~\cite{Corrigan:1979xf, Armoni:2003gp, Armoni:2003fb, Armoni:2004uu}.   The non-uniqueness of the large $N_c$ limit rests on the following simple observation.  At $N_c=3$, the fundamental (F) representation of $SU(3)_{\mathrm{color}}$ is isomorphic to the two-index antisymmetric (AS) representation, via the mapping $q^{i} \rightarrow \frac{1}{2} \epsilon^{i j k} q_{i j}$, where $q^{i}$ is a fundamental quark (F quark) and $q_{i j}$ is a two-index anti-symmetric representation quark.  Thus, when we take the large $N_c$ limit, we can take the quarks to be in the F representation;  this gives the usual `t Hooft large $N_c$ limit, which we will refer to as the large $N_c^{F}$ limit.  Alternatively, we can take the quarks to be in the AS representation, and obtain a quite different large $N_c$ limit, which we will refer to as the large $N_c^{\rm AS}$ limit.  (The phrase `large $N_c$ limit' will be used below when referring to large $N_c$ limits in general.)

In this paper we explore the question of whether large $N_c$ QCD(AS) can be in a color superconducting phase at high density, as one expects for $N_c=3$.  We focus on the high-density regime since this allows us to do controlled calculations.  While our ultimate interest is in more moderate densities, if the high-density physics at large $N_c$  is qualitatively similar to the behavior of the theory at $N_c=3$, one might hope that this persists to lower densities.  To keep the discussion concrete we will focus on  $SU(N_c)$ gauge theories  with $N_f=2$ massless flavors of Dirac fermions and compare the behavior of the case of quarks transforming either in the F representation [referred to as QCD(F)],  or in the AS representation [referred to as QCD(AS)], with both flavors transforming in the same representation of the gauge group.   One can also consider hybrid large $N_c$ limits with some flavors of quarks transforming in the F representation and others in the AS representation~\cite{Armoni:2003gp, Armoni:2003fb, Armoni:2004uu,Kiritsis:1989ge,Armoni:2005wt,Frandsen:2005mb,Sannino:2007yp,Armoni:2009zq,HoyosBadajoz:2009hb}.  We leave the exploration of the finite-density physics of such hybrid large $N_c$ limits to future work.

The large $N_c^F$ and $N_c^{\rm AS}$ limits are similar in a number of ways:  for instance, non-planar diagrams are suppressed in both, and both contain an infinite number of narrow mesons at large $N$~\cite{Armoni:2003gp, Armoni:2003fb, Armoni:2004uu}.  Furthermore, baryons appear as topologically non-trivial configurations of meson fields in both large $N_c$ limits (at least for odd $N_c$, the phenomenologically relevant case), and baryon masses scale as $N_c^1$ and $N_c^2$ respectively in the large $N_c^F$ and $N_c^{\rm AS}$ limits~\cite{Bolognesi:2006ws,Cherman:2006iy,Cherman:2006xa}.  The major difference between the two large $N_c$ limits is that quark loops are suppressed in the `t Hooft large $N_c$ limit, and are not suppressed in the large $N_c^{\rm AS}$ limit.  Thus, the dynamics and $N_c$ scaling of observables are generally quite different in the two large $N_c$ limits.

The lack of quark loop suppression in the large $N_c^{AS}$ limit is both a virtue and a vice:    Zweig's rule is behind many of the `t Hooft limit's phenomenological successes, but also behind some of its failures, such as with the $\eta'$ mass.  Which of the two large $N_c$ limits is more useful phenomenologically presumably depends on what observables one wishes to study~\cite{Armoni:2003gp, Armoni:2003fb, Armoni:2004uu,Kiritsis:1989ge,Armoni:2005wt,Frandsen:2005mb,Sannino:2007yp,Armoni:2009zq,HoyosBadajoz:2009hb,Cherman:2006xa,Cherman:2009fh}.

The large $N_c^{\rm AS}$ limit has a number of very attractive theoretical properties, chief among them being the fact that QCD(AS) in the large $N_c$ limit plays a starring role in the orientifold planar equivalence\cite{Armoni:2003gp, Armoni:2003fb, Armoni:2004uu}.  This equivalence, which holds in the large $N_c$ limit, is for the correlation functions of a broad class of charge-conjugation even operators in QCD(AS).  The equivalence implies that at large $N_c$  these these observables in QCD(AS) with $N_f$ flavors exactly coincide with the corresponding observables in $SU(N_c)$ Yang Mills theory with $N_f$ adjoint Majorana flavors.  Aside from its intrinsic interest, this observation can be used as a very useful computational tool when $N_f=1$, since YM with $N_f=1$ Majorana quarks is  nothing other than $\mathcal{N}=1$ super-Yang-Mills theory.  As a result, the orientifold equivalence has the surprising consequence that one can bring to bear all of the powerful machinery of supersymmetry to calculate some of the properties of the non-supersymmetric $N_f=1$ QCD(AS) theory \cite{Armoni:2003gp, Armoni:2003fb, Armoni:2004uu,Armoni:2005wt,Armoni:2009zq}.  Unfortunately, we do not know of a way to use the equivalence to study finite-density physics in QCD(AS), even at $N_f=1$, since the quark number current is not within the class of observables to which the equivalence applies.

Here, we examine the high-density behavior of large $N_c^{\rm AS}$ QCD in the chiral limit with a focus on the fate of the BCS and DGR instabilities, as compared to their behavior in the standard `t Hooft large $N_c$ limit.  As will be discussed below, the scale of the onset of the BCS instability in a renormalization group flow is exponentially $N_c$-suppressed in both large $N_c$ limits.  This can be traced back to the fact that the BCS condensate is not a color singlet. This fact implies that if other instabilities exist which are not as suppressed at large $N_c$, the BCS phase will not occur at large $N_c$---at least not in the asymptotically high-density regime; the phase induced by the earlier instability will prove energetically favorable.  This is precisely what happens for the case of the QCD in the large $N_c^F$ limit, where the DGR instability dominates.   However, in contrast to QCD in the large $N_c^F$ limit,  we show that the DGR instability does not occur in the large $N_c^{\rm AS}$ limit.   This raises the prospect that QCS(AS) may, indeed, have a BCS phase at large $N_c$ and high density.  Of course, if there are other instabilities toward the formation of color-singlet condensates that occur at large $N_c$ in QCD(AS), they will dominate over the BCS instability.  If such instabilities are absent, however, one might be led to hope that the color-superconducting phase may survive to large $N_c$ in QCD(AS).  Thus perhaps large $N_c$ reasoning could shed some light on finite-density phenomenology after all.

We note that the study of finite-density physics in the large $N_c^{\rm AS}$ limit of QCD was pioneered in Ref.~\cite{Frandsen:2005mb}.  However, Ref.~\cite{Frandsen:2005mb} did not fully take into account the effect of quark loops and screening.  The different behavior of quark loops is at the heart of the differences between the two large $N_c$ limits, and changes the analysis in fundamental ways, as will be clear below.

\section{Color superconductivity in QCD(AS)}
\label{sec:BCS}
At high densities, one expects quark matter to become weakly interacting, and a Fermi liquid description should become accurate.  Color superconductivity is  a consequence of the fact that  at high densities, when $\mu \gg \Lambda_{QCD}$, the quark Fermi surface is unstable due to an attractive channel in (for instance) s-wave quark-quark scattering near the Fermi surface.  The scale of the superconducting gap $\Delta$ can be approximated as the scale at which the Fermi liquid description breaks down, and it turns out  that $\Delta\sim \mu g^{-5} e^{-K/g}$, where $K$ is a numerical constant that generally depends on $N_c$, and $g$ is the Yang-Mills coupling evaluated at the scale $\mu$.  The fact that the QCD gap scales with $1/g$ rather $1/g^2$ (as is the case in conventional BCS superconductivity) is due to an infrared (IR) enhancement of the quark-quark scattering amplitudes due to long-range interactions mediated by magnetic gluons \cite{Son:1998uk}.

It is straightforward to show that the preceding discussion applies equally well to QCD(F) and QCD(AS), with the only difference between the two theories being in the constant $K$.  To see this, we briefly review the classic calculation of the color superconducting gap from renormalization group (RG) arguments as given, for instance, in Ref.~\cite{Son:1998uk}, and highlight the differences between the two large $N_c$ limits as we do so.

The basic idea of the RG approach to estimating the size of the gap is to look at the RG evolution of the quark-quark ($qq$) scattering amplitude as one integrates out degrees of freedom in successive shells around the Fermi surface \cite{Polchinski:1992ed}.   An attractive interaction (provided it is relevant or marginal) generically leads to an infrared Landau pole as one follows the RG flow toward the Fermi surface.   The Landau pole signals the breakdown of Fermi liquid theory and an instability of the Fermi surface toward the formation of a condensate, which is schematically $\langle q q \rangle \sim \Delta^3$ in the case of superconductivity.   More attractive interactions typically lead to Landau poles occurring at lower energies.  The scale of the gap associated with the formation of the condensate is then approximately given by the energy scale of the first Landau pole that occurs.

The first step in this procedure is the identification of the most attractive $qq$ interaction.  At high densities, where one is justified in only considering interactions mediated by one-gluon exchange, $qq$ interactions split into color channels, some of which are attractive and some repulsive.  The available color channels are different in QCD(F) and QCD(AS)~\cite{Frandsen:2005mb}.  For QCD(F), the available color channels are given by
\beq
\Yvcentermath1
\yng(1) \otimes \yng(1) = \yng(1,1) \oplus \yng(2) \;,
\eeq
while for QCD(AS), the color channels are given by
\beq
\Yvcentermath1
\yng(1,1) \otimes \yng(1,1) = \yng(1,1,1,1) \oplus \yng(2,1,1) \oplus \yng(2,2) \;.
\eeq
Each channel is associated with a color factor which determines the sign and magnitude of the interactions.   The totally antisymmetric channels are expected to be the most attractive ones in both cases, so that the relevant color factors are given by
\beq
C_{F} = \frac{N_C + 1}{2N_C} \quad, \quad C_{AS} = \frac{2N_C+2}{N_C}.
\eeq
For $N_c=3$, the $C_F$ color factor is the one associated with the familiar $\bar{3}$ channel, while the $C_{AS}$ color factor is associated with a channel that is only available when $N_c>3$.  Since we are interested in the large $N_c$ physics of QCD(AS), we will use $C_{AS}$ as the color factor in the analysis below, since it is associated with the most attractive channel at large $N_c$.  More generally, we denote the color factor associated the most attractive channel in a large $N_c$ limit  as $C_{R}$  where $R$ is  the representation of the fermions.

As is well known, the interaction that leads to the BCS instability is nearly-collinear s-wave $qq$ scattering.  This is because collinear scattering can be shown to be a marginal interaction, in contrast to non-collinear scattering, which turns out to be technically irrelevant near the Fermi surface~\cite{Polchinski:1992ed}.  We focus on s-wave scattering because it tends to be the most attractive channel, as well as for simplicity, although p-wave and d-wave superconductors are, in principle, also possible.

In QCD, the s-wave partial-wave amplitudes would be IR divergent were it not for screening effects.  The color-electric field is screened by the Debye mass $m_D^2 = N_f C(R)/\pi^2 g^2 \mu^2$  as can be seen by resumming quark bubble diagrams in the gluon propagator.  $C(R)$ is the Casimir coefficient of the fermion representation $R$ defined as $\delta^{ab}C(R) = \tr{\, t^{a}_{R} t^{b}_{R}}$ where the $t^{a}_{R}$ are the $SU(N_c)$ generators in the representation $R$.  It is not hard to show that  $C(\mathrm{F}) = 1/2$, $C(\mathrm{AS})= (N_c-2)/2$, so that the value of the Debye mass is representation-dependent.  The resulting differences in the $N_c$ scaling of $m_D$ can be traced back to the different $N_c$ scaling of quark loops in the two large $N_c$ limits. 

In contrast to the color-electric field, however, the color-magnetic field is not statically screened in perturbation theory: there is no perturbatively-generated magnetic Debye mass \cite{Son:1998uk}.  Instead, the magnetic field is screened dynamically due to Landau damping:  the propagator for color-magnetic gluons is given by
\beq
G(q_0,\vec{q}) = \frac{1}{q_0^2 - \vec{q}^2+\frac{\pi}{2} m_D^2 q_0/|q|}
\eeq
The dynamic damping disappears for $q_0 = 0$, but is non-zero otherwise.

An accurate estimate of the gap requires the inclusion of the effects of Landau damping: color superconductivity is due to the effects of the long-range interactions mediated by the magnetic gluons \cite{Deryagin:1992rw,Son:1998uk}.  However, a minor subtlety arises when one considers the large $N_c^{F}$ limit.  In this limit, fermion loops are suppressed, and consequently the Debye mass is suppressed by $1/N_c$. Introducing  the `t Hooft coupling $\lambda \equiv g^2 N_c$, which is held fixed in the large $N_c$ limit, the Debye mass is given by  $m_D^{F} \sim g\mu = \lambda^{1/2} \mu / N_c^{1/2}$.    Although $m_D^{F} = 0$ in the strict $N_c \rightarrow \infty$ limit, one cannot simply set $m_D^{F} = 0$ at the outset.  Doing so would lead to unscreened quark-quark scattering amplitudes, which would be IR-divergent in both the color-electric and color-magnetic channels, and the theory would be ill-defined.  Thus, $m_{D}^{F}$ can only be taken to zero (its large $N_c$ value) at the end of a problem.   In the large $N_c^{AS}$ limit, in contrast, fermion loops are not suppressed relative to gluon loops, and $m_D^{AS} \sim g N_c^{1/2} \mu =\lambda^{1/2} \mu$, so that the Debye mass is not suppressed at large $N_c$.

Given this subtlety, we keep $m_D\neq 0$ in our analysis of the RG behavior of the $qq$ interaction for both large $N_c$ limits.  It turns out, however, that the size of $m_D$ does not qualitatively change the dominant scaling of the color-superconducting gap.  The crux of the analysis is that nearly collinear $qq$ pairs generally scatter through intermediate states of energy $\delta \ll \epsilon_F$.   The RG approach consists of integrating out degrees of freedom (i.e., intermediate states) that have energies outside narrow shells around the Fermi surface $\epsilon_F \pm \delta$.  One then lets $\delta$  approach  $\epsilon_F$ while keeping track of the degrees of freedom which have been integrated out on the scattering amplitude .  This procedure yields a first-order differential equation (the RG equation) for the scattering amplitude as a function of $\delta$.   As shown in Ref.~\cite{Son:1998uk}, the RG equation for the s-wave scattering amplitude $f_s$ is
\beq
\label{eq:BCS_RG}
\frac{df_s}{dt} =  - \frac{\mu^2}{2\pi^2}f_s^2 -\frac{C_{R}}{2}\frac{g^2}{\mu^2}
\eeq
where $t = - \log( \delta/m_D)$ is the RG parameter.   The first term in the equation above is due to the `instaneous' interactions mediated by gluons that have momenta that are large enough for Landau damping to be negligible, while the second term comes from interactions mediated by softer gluons that are sensitive to the Landau damping.    It is conventional to start the RG flow at $\delta=m_D$ so that the the initial condition on $f_s(t)$ at $t=0$ is given by
\bea
f_s(0) = -C_{R} \frac{g^2}{\mu^2} \log(1/g) ,
\eea
which comes from evaluating the s-wave scattering amplitude for nearly collinear gluons in the leading log approximation.

The solution of the RG equation for small $g$ is now given by
\beq
f_s(t) \approx -\sqrt{\frac{C_R}{3}}\frac{\pi g}{\mu^2}\tan\Bigg[\sqrt{\frac{C_R}{3}}\frac{g}{2\pi}\Big(t-6\ln g\Big)\Bigg] \;.
\eeq
The amplitude hits a pole when the argument of the tangent is $\pi/2$, which lets us estimate the superconducting gap as
\beq
\Delta_{BCS} \sim m_D e^{-t} \sim \mu \sqrt{C(R)} g^{-5} \exp\Big(- \sqrt{\frac{3}{C_R}}\frac{\pi^2}{g}\Big).
\eeq
Thus we learn that the superconducting gaps in the two large $N_c$ limits can be estimated as
\bea
\Delta_{BCS}^{Fund.}  &\sim& \mu \frac{N_c^{5/2}}{\lambda^{5/2}}   \exp\Big(- \pi^2 \sqrt{\frac{6N_c}{\lambda}}\Big) \nn\\
\Delta_{BCS}^{AS}  &\sim& \mu \frac{N_c^{3}}{\lambda^{5/2}} \exp\Big(-\pi^2\sqrt{\frac{3N_c}{2\lambda}}\Big).
\eea

The gap, and thus the formation of the condensate, is suppressed exponentially in $N_c$ in \emph{both} large $N_c$ limits (our result for the large $N_c$ limit coincides with that of Son and Shuster~\cite{Shuster:1999tn}).  This is expected from general considerations, since the $\langle q q \rangle$ condensate is not a color singlet.  Note, however, that while the gap is exponentially suppressed in $N_c$ in both large $N_c$ limits, the gap is actually exponentially larger in the large $N_c^{\rm AS}$ limit.

The exponential suppression in $N_c$ of the color-superconducting gap implies that if there are {\it any} other condensates possible that are \emph{not} exponentially suppressed in the large $N_c$ limit, the color-superconducting instability will  not  occur via the mechanism sketched above.   The key point is that the phase associated with the non-suppressed condensate will be exponentially favorable energetically compared to the color superconducting phase.

 In the large $N_c^{F}$ limit, it is known that a DGR instability  of the Fermi surface due to the formation of a color-singlet quark-antiquark condensate occurs \cite{Deryagin:1992rw,Shuster:1999tn}.  This means that in the large $N_c^{F}$ limit, the preceding argument for color superconductivity does not apply, and the ground state of quark matter is not expected to be a color superconductor.  It is either a chiral-density wave or some lower-energy configuration.  As we will see in the next section, however, the DGR instability does not occur in the large $N_c^{AS}$ limit, and it is not known whether {\it any}  color singlet condensates can actually form.

\section{DGR instability}\label{sec:DGR}
One of the alternatives to the $\langle q q \rangle$ condensate associated with color superconductivity is a $\langle \bar{q} q\rangle$ condensate first identified by DGR \cite{Deryagin:1992rw}.  This condensate, which is associated with the formation of chiral-density waves,  is a position-dependent color singlet
\beq
\langle \bar{\psi(x)} \psi(y) \rangle = e^{i \vec{P} \cdot (x+y)} \int d^4q e^{-iq(x-y)}f(q),
\eeq
where $\vec{P}$ is a vector satisfying  $|\vec{P}|=\mu$.   If the Fermi surface has an instability with respect to the formation of this, condensate quark matter would presumably be in a chiral-density wave phase, unless there exists an energetically cheaper phase.  At large $N_c$, if this condensate can form, it would beat the $\langle q q \rangle$ condensate associated with color superconductivity, since the $\langle \bar{q}q\rangle$ condensate is not exponentially suppressed by $1/N_c$.    It is known that this is exactly what prevents color superconductivity in the large $N_c^F$ limit:  the DGR instability of the Fermi surface occurs at large $N_c$ when the quarks transform in the fundamental representation of $SU(N_c)$.  In this section, we will review the conditions necessary for the occurrence of the DGR instability using an RG approach first developed by Son and Shuster, and show that the instability does \emph{not} occur in the large $N_c^{AS}$ limit.  The key role will be played by the different $N_c$ scaling of the Debye mass $m_D$ in the two large $N_c$ limits and the effects of the Debye mass on the RG flow.

Paralleling the arguments in the preceding section, we first identify the most attractive channel for the $\bar{q}q$ interaction.  This step is trivial in this case, since a color-singlet channel is available for $\bar{q} \otimes q$ in both QCD(F) and QCD(AS), and this will clearly be the most attractive channel at large $N_c$.  Next, we determine an RG equation for the $\bar{q}q$ scattering near the Fermi surface at high density, where asymptotic freedom implies that one-gluon exchange makes the dominant contribution to the interactions.  Finally, we estimate the scale of the breakdown of the Fermi liquid description due to a Landau pole, and thus estimate the scale of the DGR gap $\Delta_{\rm DGR}$.

%%%%%%%%%%Figure:Ladder %%%%%%%%%%%%%%%%%%%%%%%%%
\begin{figure}[t]
\centering
\includegraphics[width=0.4 \textwidth]{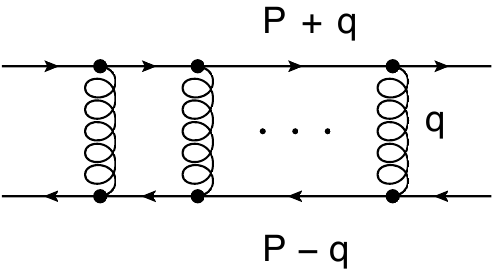}
\caption{The dominant contribution to the $\bar{q}q$ scattering amplitude for $\mu \gg \Lambda_{\rm QCD}$ is from ladder diagrams, and a typical diagram is pictured above.  The momentum flow through one of the rungs of the ladder is shown;  as is discussed in the text in a particular kinematic regime the amplitude looks like that of a 2D theory.}
\label{fig:Ladders}
\end{figure}
%%%%%%%%%%%%%%%%%%%%%%%%%%%%%%%%%%%

Recalling that we are working at high densities where $\lambda(\mu) \ll 1$, it is not hard to see that the dominant contribution to the $\bar{q}q$ scattering amplitude will come from ladder diagrams of the sort shown in Fig.~\ref{fig:Ladders}.  Each rung in the ladder diagram makes a contribution to the scattering amplitude that can be schematically (neglecting Lorentz, spin, and flavor indices) written as
\beq
\label{eq:ladder_loop}
\int{ \frac{d^4q}{(2\pi)^4} F(\vec{P}+\vec{q}) G(q) F(\vec{P}-\vec{q}) } ,
\eeq
where $F$ and $G$ refer to the fermion and gluon propagators, respectively.  To understand the RG flow of the $\bar{q}q$, it is easiest to proceed as done by Son and Shuster in their seminal paper.  The key observations is that that near the Fermi surface, in a certain kinematical region (which makes the dominant contribution to the RG flow), the integral above simplifies and becomes effectively two-dimensional, and the $\bar{q}q$ scattering amplitude associated with the ladder diagrams is then describable by a two-dimensional theory. Since it turns out that the dominant part of the RG flow takes place precisely in this kinematic region, one can determine the RG behavior of the $\bar{q}q$ scattering amplitude by studying the RG behavior of the 2D effective theory.

To see how this works, note that near the Fermi surface, it is useful to write $\vec{q}$ as a sum of two vectors, one parallel to $\vec{P}$, with magnitude $q_{||}$, and one perpendicular to $\vec{P}$, with magnitude $q_{\perp}$.  We assume that $q_{||}, q_{\perp} \ll \mu$ since we are interested in the kinematics near the Fermi surface.  We suppose that for all the internal fermion lines in the ladder diagram, $q_{\perp} \sim \delta$, where $\delta$ is an arbitrary momentum scale.  (When the system has a DGR instability, it turns out that $\delta \sim \Delta_{\rm DGR}$).

Let us now determine the kinematic regime in which the ladder diagrams are describable in terms of a two-dimensional theory, which will have $q_0$ and $q_{||}$ as its degrees of freedom.  In terms of $q_{||}$ and $q_{\perp}$, the fermion propagator looks like
\beq
F^{-1}(q) \sim i q_0 + | \vec{P} + \vec{q}| - \mu \approx i q_0 + q_{||} + \frac{q^2_{\perp}}{2\mu} .
\eeq
If $q_{||}\gg q_{\perp}^2 /\mu \sim  \delta^2 / \mu$, the fermion propagators do not depend on $q_{\perp}$.  In this regime, the only dependence on $q_{\perp}$ is in the gluon propagator, which looks like
\bea
G^{E}(q)^{-1} &\sim& q_0^2 + q_{||}^2+q_{\perp}^{2} + m_D^2 \\
G^{M}(q)^{-1} &\sim& q_0^2 + q_{||}^2+q_{\perp}^{2}  +\frac{\pi}{2} m_D^2 q_0/|q| \nnb
\eea
for electric and magnetic gluons.  When $q_{||}\sim q_{0} \ll  \delta$, the integral over $q_{\perp}$ simplifies to a logarithmic one, which is cut off in the UV by $\Delta$ and in the IR by the largest of $q_{||}$, and $m_{D}$ for electric gluons, and the largest of $q_{||}$ and $m_{D}^{2/3} q_{||}^{1/3}$ for magnetic gluons.  Thus when $ \delta^{2}/\mu \ll q_{||} \ll \delta$, the entire effect of the integration over $q_{\perp}$ in each rung of the ladder diagrams can be encoded by factors of
\bea
\label{effective_couplings}
&&\textrm{Electric gluons}:  \lambda_0(q_{||})  \equiv \frac{g^2}{4\pi} \ln \left(\frac{\delta}{\max(q_{||},m_{D})}\right) \\
&&\textrm{Magnetic gluons}:  \lambda_1(q_{||})  \equiv \frac{g^2}{4\pi} \ln\left(\frac{\delta}{\max(q_{||},m_{D}^{2/3} q_{||}^{1/3})}\right) \nnb
\eea

Thus we see that when $\delta  \gg  q_{||} \gg \delta^{2}/\mu$, the ladder diagrams can be described by an effective two-dimensional theory of fermions where the effects of the gluons are encoded in scale-dependent couplings for electric and magnetic interactions which are given by the factors in Eq.~(\ref{effective_couplings}).  When $q_{||}$ falls outside this range, the RG flow is negligible, as is explained in Ref.~\cite{Shuster:1999tn}.  Thus, for the DGR instability to occur, there must be an IR Landau pole within the range $\delta  \gg  q_{||} \gg \delta^2/\mu$.  To proceed with the analysis, let us write the lagrangian describing the 2D effective theory, which can be written in terms as a doublet of 2D Dirac fermions associated with particles with momenta near $\vec{P}$ and $-\vec{P}$ (the antiparticles are heavy and decouple from the effective field theory).  A 4D Dirac fermion can be written (in the chiral basis) as $\psi^{T} = (\psi_{L1}, \psi_{L2}, \psi_{R1}, \psi_{R2})$.  The 2D Dirac fermions can be written as
\beq
\varphi= \left(\begin{array}{c} e^{-i \mu z} \psi_{L2} \\ e^{i \mu z} \psi_{R2} \end{array}\right), \;\;
\chi = \left(\begin{array}{c} e^{-i \mu z} \psi_{R1} \\ e^{i \mu z} \psi_{L1} \end{array}\right) ,
\eeq
where we have assumed for definiteness that $\vec{P}$ is pointing in the $z$ direction.  In terms of the doublet $\Psi^{T} = (\varphi, \chi)$, the 2D effective theory is described by the Lagrangian
\beq
\mathcal{L} = \bar{\Psi} \slashed{\partial} \Psi - \lambda_0(q_{||}) \left(\bar{\Psi} \gamma^{0} \frac{T^{a}}{2} \Psi \right)^2
+ \lambda_1(q_{||}) \left(\bar{\Psi} \gamma^{1} \frac{T^{a}}{2} \Psi \right)^2 \, .
\eeq

The 2D effective theory is a Thirring-like model with different couplings for electric and magnetic interactions.  The RG equations for $\lambda_0, \lambda_1$ can be found from looking at one-loop diagrams for the 2D fermions in the usual way, and one can show that the RG equations for $\lambda_{+} \equiv (\lambda_0+\lambda_1)/2$ and $\lambda_{-} \equiv (\lambda_0 - \lambda_1)/2$ decouple:
\bea
\label{DGR_RG}
\frac{\partial \lambda_{+}(s,u) }{\partial s} &=& \frac{N_c}{\pi} \lambda^{2}_{+} (s,s) \, ;\\
\frac{\partial \lambda_{-}(s,u) }{\partial s} &=& 0 \, , \nnb
\eea
where $s$ is the RG parameter, and $u \equiv \ln(\delta/q_{||})$.  The initial conditions for the RG equations will have to be given in piecewise form depending on the relative size of $q_{||}$ and $m_D$.  For the initial conditions, we have
\bea
q_{||}<m_D &:&\; \lambda_{+}(0, u) = \frac{g^2}{4\pi}\left( \frac{5}{6} \ln \left(\frac{\delta}{q_{||}}\right) +\frac{1}{6} u\right) \, ; \nnb \\
q_{||}>m_D &:&\; \lambda_{+}(0, u) =  \frac{g^2}{4\pi} u \, .
\eea
It can be shown that the solution of Eq.~\ref{DGR_RG} with the above boundary conditions is
\beq
\label{RG_sol_noMD}
 \lambda_{+}(s, u) =   \frac{g^2 N_c }{4\pi^2}  \tan\left(   \frac{g^2 N_c }{4\pi^2}  s \right)  +   \frac{g^2}{4\pi} (u-s)
\eeq
for $s<\ln\left(\delta/m_D\right)$, and
\beq
\label{RG_sol_withMD}
 \lambda_{+}(s, u) = \frac{ g^2 N_c}{4\pi^2\sqrt{6}} \tan\left( \frac{g^2 N_c }{4\pi^2\sqrt{6}} (s+c) \right) + \frac{g^2}{4\pi} (u-s)
\eeq
for $s>\ln\left(\delta/m_D\right)$, where the constant $c$ is determined by demanding continuity at $s=\ln\left(\delta/m_D\right)$, yielding
\beq
c= \frac{ g^2 N_c \sqrt{6}}{4\pi^2} \arctan\left(\sqrt{6} \tan \left( \frac{g^2 N_c }{4\pi^2} \ln \left(\delta/ m_D\right)\right) \right) .
\eeq

Suppose first that $m_D < \delta$.  Then $\lambda_{+}$ as given by Eq.~(\ref{RG_sol_noMD}) is the relevant solution, and $\lambda_{+}$ hits a Landau pole when the argument of the tangent reaches $-\pi/2$, so that we can estimate the size of the DGR gap as
\beq
\Delta_{DGR} \sim \delta e^{-\frac{2\pi^3}{g^2 N_c}}.
\eeq
Since the RG evolution only takes place when $\delta^2/\mu \ll q_{||}$, we must have $\Delta_{DGR} \gtrsim \delta^2/\mu$ if the Landau pole is to be reached during the RG evolution.  When this relation is saturated, $\delta \sim \mu \exp(-\frac{2\pi^3}{g^2 N_c})$.  At this point $\Delta_{DGR} \sim \delta^2/\mu$, so we obtain the estimate
\beq
\Delta_{\mathrm{DGR}} \sim \mu e^{-\frac{4\pi^3}{g^2 N_c}}, \; m_D<\Delta_{\mathrm{DGR}} .
\eeq

Note that the size of the  DGR gap depends on the `t Hooft coupling $\lambda = g^2 N_c$ rather than just on $g$ (as was the case for the color superconducting gap), so that the DGR gap is much larger than the color superconducting gap at large $N_c$, provided that $m_D < \Delta_{DGR}$.  Since $m_D \sim \mathcal{O}(N_c^{-1/2})$ in the large $N_c^{F}$ limit,  we see that the DGR instability beats the BCS instability in this limit.  Accordingly  the ground state of high-density quark matter looks very different in the large $N_c^F$ world compared to the $N_c=3$ world.

The situation is profoundly different in the large $N_c^{AS}$ limit, since $m_D^{AS} \sim \mu g N_c^{1/2} \sim \mu \lambda^{1/2}$:  quark loops are not suppressed and the gluons are screened by the quarks.  Accordingly it is not consistent to have $m_D <\delta$.  To proceed we must consider what happens when $m_D>\delta$.  The coupling then runs as in Eq.~(\ref{RG_sol_withMD}), and $m_D$ now {\it does} affect the RG flow.  Indeed, $m_D$ is large enough to essentially cut off the RG flow before  a Landau pole is reached, thereby preventing the DGR instability.

 To see how this happens, note that the coupling in Eq.~(\ref{RG_sol_withMD}) hits a Landau pole at the energy scale $E_L$ when the argument in the tangent hits $\pi/2$, so that
\bea
\ln\left(\frac{E_L}{\delta}\right) &\sim& \frac{ g^2 N_c \sqrt{6}}{4\pi^2} \tan^{-1}\left(\frac{1}{\sqrt{6}} \cot \left( \frac{g^2 N_c }{4\pi^2} \ln{\frac{\delta}{m_D}}\right) \right) \nnb\\
 &+&  \ln{\frac{\delta}{m_D}}.
\eea
We must again impose the condition that $\delta^2 / 2\mu \lesssim E_L$, since outside this region there is no RG flow and any Landau poles outside this region would be unphysical.  Some algebra then shows that this condition on $E_L$ implies that we must have $m_D < m_D^{\mathrm{critical}}$, where
\bea
m_D^{\mathrm{critical}}&\lesssim& \mu \exp\left(-\frac{4\pi^2 c_0}{g^2 N_c}\right), \;\; \rm{with} \\
c_0 &=& \sqrt{6} \tan^{-1}{\frac{1}{2}}+2\tan^{-1} {\sqrt{\frac{2}{3}}}
\eea
if the condition $\delta^2 / 2\mu \lesssim E_L$ is to be met.  Since $m_D^{AS} \sim \mu g N_c^{1/2} > m_D^{\mathrm{critical}}$, the DGR instability does not occur in the large $N_c^{AS}$ limit.

\section{Discussion}

As we have seen, the large $N_c^{F}$ and large $N_c^{AS}$ worlds appear to be quite  different at finite density.  Before discussing these differences in detail, it is useful to briefly discuss a fundamental imitation  of the type of analysis done here.  Note that the ultimate question under consideration is the lowest energy state of bulk matter at finite density.  Thus we are looking for a global minimum.

However, the techniques we used are local.  One starts with a Fermi liquid type description, assumed to be reasonable at high densities where the quarks are weakly coupled, and looks for {\it local} instabilities toward the formation of various types of condensates.  If such instability can be shown to exist in a regime where one can calculate reliably, one  knows that the ground state is {\it not} a Fermi liquid.  One does not necessarily know the actual ground state.

There are two potential difficulties here.  Firstly we can only get insight into states which are in some sense ``close'' to a Fermi liquid.  One can only do reliable calculations to show local instabilities to the extent that the relevant dynamics involve modes very near the would-be Fermi surface.  However,  if a global minimum which is {\it not} associated with a local instability were to exist, these methods would not be sensitive to it.    The second difficulty is simply that the number of possible condensates is quite large once one includes color, flavor, Dirac and spatial degrees of freedom.   Thus, even if one has found an instability leading to the formation of a condensate, one cannot be certain it is the lowest energy phase unless one has explicitly checked that all  other possible condensates either do not form, or are associated with phases  of higher energy.

This issue is quite generic and does not depend on the large $N_c$ limit.  Thus, it is fair to say that in some deep epistemological sense we do not really {\it know}  that the ground state of QCD with $N_c=3$ is a color superconductor even at asymptotically high densities.  However, in this case there are very good reasons to believe that the ground state really is a color superconductor.   This system has been extensively studied for a long time and no plausible alternative ground state has been found.  Moreover, during this study considerable intuition has built up about the system and it seems implausible that a lower ground state exists.  The situation is different for QCD in either of the large $N_c$ limits considered here.  These systems have not been extensively studied, so intuition about these systems is lacking.  Moreover, the large $N_c$ limit at finite density can be quite subtle and act in surprising ways \cite{Cohen:2004mw,Bringoltz:2009ym}.

What do we know about these systems? For the case of QCD in  the large $N_c^{F}$ limit at high density, we know there is a DGR instability as well as a BCS type instability.  The existence of these instabilities means that we know that the ground state is {\it not} a Fermi liquid.  Furthermore, since the DGR instability is exponentially stronger than the color superconducting one, we also know that the standard color superconducting state is \emph{not} the ground state.  We do not know with certainty that the ground state of the system is a chiral-density wave phase, but it is the best candidate known for the ground state. It is not implausible that the actual ground state is indeed the DGR phase in the large $N_c^{\rm F}$ limit.

The case of QCD in the high-density large $N_c^{AS}$ world is quite different.  In this world the DGR instability is absent, and thus we know that  a simple  DGR chiral-density wave is \emph{not} the ground state.  The difference between the behavior in the two large $N_c$ limits can be traced back to the lack of suppression of quark loops in QCD(AS) at large $N_c$.  This means  that gluons are screened with different strengths in the two large $N_c$ limits:  screening is suppressed with $N_c$ in QCD(F), but is unsuppressed in QCD(AS).  This difference turns out not to qualitatively affect the estimate of the scale of the color-superconducting gap, but has a dramatic effect on whether the DGR instability occurs.

What, then, is the ground state of high-density quark matter in the large $N_c^{\rm AS}$ limit?  Our present state of knowledge is consistent with it being a color superconductor.  Given the caveats above, we cannot say with certainty that it is actually so.  It may seem that this caution is a mere quasi-philosophical quibble, as with the case of color superconductivity at $N_c=3$.  However, this is not the case.  QCD in the large $N_c^{\rm AS}$ limit is rather different from the $N_c=3$ case in one critical way: the gap is exponentially suppressed in $N_c$.  Thus, the existence of {\it any} instability towards the formation of a condensate  which is color singlet --- and thus \emph{not} exponentially suppressed in $N_c$  --- is expected to lead to a phase which is energetically favorable compared to the color superconducting phase.  The reason is simply that  the associated gaps would presumably scale with the `t Hooft coupling $\lambda$ rather than with the Yang-Mills coupling $g$.   

In this work, we investigated one such possible color-singlet condensate (the DGR condensate), and found that the effects of screening in large $N_c^{\rm AS}$ QCD prevent it from forming.  One might, of course, wonder whether there are any other instabilities of Fermi surface toward the formation of other color singlet condensates.  For instance, one might worry about a ferromagnetic condensate with the schematic form $\langle \bar{q} \gamma^{\mu} \gamma^{5} q \rangle$~\cite{Tatsumi:1999ab,Nakano:2003rd,Ohnishi:2006hs}.   However, this condensate does not form in the chiral limit where $m_q=0$~\cite{Nakano:2003rd}, so it would not compete with the color superconducting condensate in the large $N_c^{\rm AS}$ limit with massless quarks.  An investigation of other possible condensates in QCD(AS) at large $N_c$ is an interesting subject for future work.

However, it is certainly possible that no instability towards a color singlet condensate exists.  If this is true, then the large $N_c^{\rm AS}$ world is qualitatively close to the $N_c=3$ world as far as high-density physics is concerned, at least in the sense that the the ground state might be a color superconductor.

In the optimistic scenario where one would suppose that the ground state of  high-density QCD in large $N_c^{\rm AS}$ limit is indeed a color superconductor, it is not immediately clear whether the resulting large $N_c^{\rm AS}$ world is close enough to the $N_c=3$ world to make large $N_c^{\rm AS}$ investigations phenomenologically useful.  There are reasons for caution. At low densities ($\mu \ll \Lambda_{QCD}$) the large $N_c$ world looks quite different from the $N_c=3$ one:  in  the large $N_c^{\rm AS}$ limit (just as in the large $N_c^{\rm F}$ limit), baryons are thought to form Skyrme crystals~\cite{Klebanov:1985qi,McLerran:2007qj}, which does match the known phenomenology for $N_c=3$.  However, it is not known what the large $N_c^{\rm AS}$ world looks like at higher densities, and if it indeed looks qualitatively similar to the $N_c=3$ world at asymptotically high densities, it is certainly worth investigating whether the large $N_c^{AS}$ limit of QCD might be useful when the chemical potential is in the phenomenologically interesting range $\mu/\Lambda_{\rm QCD} = \mathcal{O}(1)$.

{\it Acknowledgements.}  A.~C. thanks the Nuclear Theory group at the Tokyo Institute of Technology for its very warm hospitality during the initial stages of this work, and we thank Mark Alford, Adi Armoni, Paulo Bedaque and Makoto Oka for illuminating discussions.  The support of the US DOE under grant DE-FG02-93ER-40762 is gratefully acknowledged.

\end{document}